\documentclass[psfig,10pt,epsf,amssymb]{article}

\usepackage{graphicx}

\newcommand{\be}{\begin{equation}}
\newcommand{\ee}{\end{equation}}
\newcommand{\bea}{\begin{eqnarray}}
\newcommand{\eea}{\end{eqnarray}}

\newcommand{\half}{\frac{1}{2}}

\def \square {\hbox{$\sqcup\!\!\!\!\sqcap$}}

\begin{document}
\title{Analog Models beyond Kinematics }
\maketitle
\begin{center}
\author{ {\bf Serena Fagnocchi}\\
\footnotesize \noindent{\it Centro  Studi e Ricerche "Enrico Fermi", Compendio Viminale, 00184 Roma, Italy\\
INFN sezione di Bologna, via Irnerio 46, 40126 Bologna, Italy\\}
e-mail: fagnocchi@bo.infn.it}
\end{center}

\maketitle

\begin{abstract}
In this paper I discuss the extension of the analogy between gravitation and some systems of condensed matter physics from kinematics to dynamics.  I will focus my attention on  two applications of the analogy to the dynamics of fluids that have been recently proposed: the study of backreaction effects and the calculation of the depletion in Bose-Einstein condensates, showing how this extension is possible and stressing the main differences with respect to the gravitational context. I will conclude with some remarks about the actual reliability of the proposed scheme, pointing out the basis issues that have still to be addressed.
\end{abstract}

One of the most peculiar features of actual theoretical physics is that some of its branches  suffer an incredible lack of experimental data that seriously risks to compromise their development. Among them is Quantum Field Theory (QFT) in curved space, with its most famous result -- the Hawking effect \cite{hawking}, widely considered a milestone of theoretical physics -- still unobserved. In contrast with what previously believed about black holes, Hawking, studying the propagation of quantum fields on the background of a collapsing body forming a black hole\footnote{This effect has been obtained within the semiclassical  theory of Gravity, where the spacetime is described by the classical Einstein theory of General Relativity (GR), while the fields propagating on this background are quantized.}, showed that a thermal flux of particles emerges after the horizon forms, and therefore the black hole radiates. It is of fundamental importance  that it is a pure kinematical effect, related only to the geometrical propagation of the field modes next to the horizon and completely independent of the details of the collapse itself. Therefore it will occur whenever the same spacetime structure emerges.\\
As already pointed out, for the extreme weakness of the produced effect, the Hawking radiation has so far no experimental observation confirming it: in fact for astrophysical black holes the temperature associated to the Hawking emission is of order $10^{-8}\ K$ or lower, far beyond the CMB temperature ($\sim 3\ K$) which hoplessly covers it. A possible test in next generation accelerators collisions has  been also proposed, provided an  ad hoc extra-dimensions theory lowers the scale of Quantum Gravity at $\sim TeV$ \cite{AA}.\\ A more promising way out can be in  the so called {\it Analog Models}, i.e. some systems of condensed matter which, under some circumstances, behave as gravitational ones. As shown by Unruh in his seminal work \cite{unruh81}, the fluctuations in an Eulerian fluid propagate {\it exactly} as a massless scalar field on a fictitious curved  background, whose metric tensor depends on the characteristics of the underlying fluid. In fact  an inviscid irrotational fluid is described by the hydrodynamical equations
\be \label{hydroeqs}\partial_t\rho + \vec \nabla
\cdot (\rho\vec v)=0 \, ,\qquad\qquad \partial_t \theta + \frac{1}{2}{\vec v}^2 +\mu(\rho)=0 \, , \ee with $\rho$ the density, $\vec v={\vec \nabla}\theta$ the velocity,
$\mu(\rho)\equiv \frac{du}{d\rho}$ and $u$ the internal energy. 
Expanding eqs.(\ref{hydroeqs}) at first order in the linear fluctuations $\rho_1$ and $\theta_1$ around a mean field solution of eqs.(\ref{hydroeqs}) one gets
\be\label{box}
\square \theta_1=0\, ,\qquad\qquad
\rho_1=-\frac{\rho}{c^2}\left(\partial_t \theta_1+\vec \nabla \theta \cdot  \vec \nabla  \theta_1  \right)\, ,
\ee where $\square \equiv\frac{1}{\sqrt{-g}}\partial_\mu(\sqrt{-g}g^{\mu\nu}\partial_\nu)$
is the covariant d'Alembertian calculated from the so called "acoustic metric"
\be\label{gmunu} g_{\mu\nu}\equiv\frac{\rho}{c}\left(
\begin{array}{cc}
-(c^2 -v^2)& -\vec v^T\\
-\vec v& \mathbf{1}
\end{array}\right)\ , \ee and $c$ is the sound speed $c^2=\rho \left. \frac{d\mu}{d\rho}\right|_{\rho_0}$. \\
Thus the field $\theta_1$ satisfies the equation of motion for a massless scalar field in the curved background described by the tensor metric (\ref{gmunu}). For a homogeneous fluid the acoustic metric reduces to the flat minkowskian one, while it can show a black hole form if $|\vec v|=c$: there $g_{00}$ vanishes and a sonic horizon forms \cite{reviewvisser}. The supersonic region is the acoustic analog of the black hole, where sound is trapped as light is in the gravitational field of a black hole.\\
Once the hydrodynamical fluctuations are quantized, following Hawking's arguments, a flux of phonons (the sonic quanta) at a temperature $T_H=\frac{\hbar k}{2\pi \kappa_B c}$ is expected, where $k=\half \nabla(c^2-v^2)|_{hor}$ is the acoustic analog of the surface gravity of the horizon \cite{unruh81}. This is the sonic analog of the Hawking effect. For this reason hydrodynamical systems can in principle be used as "laboratories" to observe this effect (among other effects otherwise unfathomable). One of the most promising setting to this aim are Bose Einstein condensates (BEC), for which $T_H\approx 10 \ nK $ with respect to a condensate temperature $T_c\approx 100\ nK$ \cite{stime}. Unfortunately, the analog Hawking radiation is still difficult to point out directly, both for the fact that $T_H$ is actually very low and for the noise produced by other effects that risks to  cover it. Yet the recent technological improvement in handling BEC and the possibility to study analog models more in depth make us confident that it can be soon observed.

\section*{Dynamics}
As shown in the previous section, the analogy works perfectly at the kinematical level: it occurs for the amazing fact that sound waves in a fluid "feel" inhomogeneities exactly as a quantum field on a gravitational background "feels" curvature and it is therefore possible to use the  GR formalism to describe their motion.
But the features of the acoustic metric strictly depend on the underlying medium and its flow, which can be described in no way using the GR framework: they are indeed non-relativistic ($c_{sound}\ll c_{light}$) and newtonian.
The equations governing the dynamics in the two setting evidently differ: on one side are Einstein equations, on the other hydrodynamical equations. \\Does this mean a full stop in the application of analogy beyond the kinematical issues?  Or is there still the possibility to go further  addressing also dynamical ones?\\
I will show how is still possible, with special care, to use analog models to show up also behaviors depending on the dynamics. Of course the analogy will be not "perfect" any more and the outcomes in the condensed matter context could have nothing to do with what known in GR. Yet the formalism developed in QFT in curved space turns out to be a powerful tool to evaluate analytically quantities otherwise nasty to evaluate with the standard techniques.\\ As examples, I will show in the following two seemingly very different applications of this formalism: the backreaction of the  analog Hawking radiation and the depletion in BEC.

\subsection*{Backreaction in sonic black holes}
As already mentioned, Hawking radiation is still difficult to be directly observed even in the acoustic setting. It could be easier to try and see it indirectly through the modifications its presence will induce on the fluid: this beck-effect is the so called backreaction. Of course it will depend on dynamics and there is no reason to expect that what found for gravitational black holes still holds for these completely different objects. \\ In GR backreaction is governed by the semiclassical Einstein equations, namely $G_{\mu\nu}(g_{\alpha\beta})=\hbar\langle T_{\mu\nu}(g_{\alpha\beta})\rangle$, where  $\langle T_{\mu\nu}\rangle$ is the stress energy tensor of the quantum field which acts as a source in the semiclassical equations and modifies the background spacetime. The first issue is to find the analog of the semiclassical equations for the hydrodynamical systems. In \cite{backreaction} it is shown how to  solve this problem going further in the expansions of eqs.(\ref{hydroeqs}) up to quadratic order in the fluctuations, yielding
\be  \partial_t\rho_B + \vec\nabla \cdot (\rho_B\vec v_B)
=-\partial_{i} (
\sqrt{-g^B} \langle T^{0i}\rangle)\ ,\qquad
\dot \psi_B + \frac{1}{2}\vec v_1^2 + \mu(\rho_B)
=\frac{1}{2}\frac{\rho_B}{c} \langle T \rangle\ ,
\label{backeqs}\ee with the subscript $_B$ meaning "corrected by backreaction".\\ In eqs.(\ref{backeqs}) the pseudo-stress energy tensor for the fluctuations (namely for a massless scalar field $\theta_1$) has been introduced: $T_{\mu\nu}=\partial_\mu \theta_1\partial_\nu\theta_1-\half g_{\mu\nu}\partial_\alpha\theta_1\partial^\alpha\theta_1$ and $T$ is its trace.
This procedure is equivalent to what one usually do in QFT in curved space to write down the backreaction equations for gravitational waves. It is worth noting that eqs.(\ref{backeqs}) are the most general equations governing backreaction effects in the hydrodynamical setting for $c=const$: they are not linked to the presence  of a sonic black hole and describe classical as well as quantum fluctuations backreaction. All the informations relative to the different situations are enclosed in $\langle T_{\mu\nu}\rangle$: the brackets are  stochastic mean values in describing classical fluctuations or expectation values in the quantum case. The quantum state which properly describes black hole formations and Hawking radiation is the so called Unruh state \cite{frolov}. \\
Under some simplifying hypothesis,  eqs.(\ref{backeqs}) has been solved analytically  \cite{backreaction} and numerically \cite{nagar} for a sonic black hole in the Unruh state. It has been shown that, due to the presence of the Hawking radiation, next to the horizon the fluid slows down and the black hole size shrinks. Moreover the emission temperature gets lower and lower as the evaporation goes on, reaching zero temperature emission in an infinite amount of time. This is not a surprising behavior: the evaporation proceeds at the expenses of the kinetic energy of the fluid, which in fact slows down, and the temperature (related to the velocity on the horizon) 
also decreases. Yet it is completely different from what is known about Schwarzschild black holes, that get hotter and hotter as the evaporation goes on. The sonic black hole  looks rather like a near-extremal charged black hole: this kind of black hole reaches zero temperature emission in an infinite time, but with different power law with respect to the sonic ones \cite{review}.

\subsection*{Depletion in Bose-Einstein condensates}
As well as other ultra-cold atoms systems, BEC can be described in some regimes by hydrodynamical equations \cite{stringari}. 
In the mean field limit the bosonic operator $\hat \Psi$ describing the system can be split as $\hat \Psi=\psi +\hat\varphi$: $\psi=\langle \hat\Psi\rangle$  describes the condensed part, while $\hat \varphi$ is a quantum fluctuation. Under some hypothesis and writing $\psi=\sqrt{\rho}e^{i\theta/\hbar}$, the condensed part satisfies the classical equation of motion analog to eqs.(\ref{hydroeqs}), where now $\rho$ is the density of particles and $\vec v=m^{-1}\vec\nabla \theta$ the velocity of the fluid.  In practice this description is possible if the scale of variation of the inhomogeneities $L$ is much bigger than the healing length $\xi=\hbar/(\sqrt{2}mc)$, that is related to the mean distance between the constituents of the fluid. Moving to the fluctuation part, without loss of generality, $\hat \varphi$ can be rewritten in the acoustic representation as \be
\label{acoustic.repr}
\hat \varphi=e^{i\theta/\hbar}\left( \frac{1}{2\sqrt{\rho}}\hat \rho_1+i\frac{\sqrt{\rho}}{\hbar}\hat \theta_1\right)\, ,
\ee  with $\hat \rho_1$ and $\hat \theta_1$ two real fields. The equations describing them assume the hydrodynamical form and can therefore be rewritten using the gravitational formalism as in eqs.(\ref{box}) \cite{reviewvisser}. It is worth stressing that this description is correct only for modes with wavelength $\lambda\gg \xi$: the modes with very high frequency in fact do not obey the relativistic dispersion relation  $\omega=ck$ implicit in the GR framework, but the full Bogoliubov dispersion relation for BEC, namely $\omega=ck\sqrt{1+\xi^2k^2}$.
\\ In the hydrodynamical regime it is therefore possible to use BEC as labs for testing GR effects, but also to handle the analogy to give analytical estimations of quantities that are hardly evaluable with the standard tools. In this spirit, QFT in curved space has been used to work out the depletion for a non-homogeneous BEC \cite{bec}. The total particles density can be written as a sum of condensed and non condensed part
\be
\rho_{tot}=\langle \hat \Psi^\dagger \hat \Psi\rangle=\psi^*\psi+\langle \hat \varphi^\dagger \hat \varphi\rangle=\rho+\tilde\rho \, .
\ee The last term describes the so called depletion: because of quantum and temperature fluctuations, a non vanishing fraction of the total particles number is not in the
condensed phase. The depletion is explicitely known only for homogeneous BEC where $\rho=\rho_0=const$. At low temperature ($\kappa_B T\ll \rho_0 g$), it reads \cite{stringari}
\be\label{depletion.q+t}
\tilde \rho_0=
\frac{8}{3\pi^{1/2}} (\rho_0 a)^{3/2}\left[1+\left(\frac{\pi \kappa_B T}{2 \rho_0 g}\right)^2 \right]\, .
\ee
Moving to non homogeneous condensates with vanishing velocity, the depletion is usually approximated by eq.(\ref{depletion.q+t}) with now $\rho_0$ replaced by the actual space-time varying density $\rho(\vec x, t)$, in the so called {\it local density approximation} (LDA) \cite{huang}.\\ It is clear that this approximation is not able to take the effects of spatial derivatives into account. 
It is worth noting that the derivation of the zero temperature quantum depletion, which does not vanish in the $T\rightarrow 0$ limit,  is based on the full Bogoliubov dispersion relation for the modes.
Unlike this, the low temperature correction comes from
the phononic part of the spectrum only, i.e. that one assumed in the GR analogy. The integral giving the thermal depletion is dominated by modes with wavelength $\lambda \gg \xi$, so the result is insensitive to the high $k$ behavior of the dispersion relation. Therefore, while the correct description of the ground state requires the knowledge of the full "microscopic" theory\footnote{The linear phonon dispersion relation, characteristic of a massless relativistic free field would lead (after renormalization) to a vanishing quantum depletion.}, deviations from the ground state value can be evaluated using the low energy "macroscopic" phonon theory. In other words, we assume that the corrections coming from the inhomogeneities of the condensate to the ground state result, sensitive to the whole quantum theory describing the system, are determined by those modes with wavelength comparable with the scale of the inhomogeneities (as curvature is seen by modes with $\lambda$ of the same order of the curvature itself). Those modes therefore satisfy the hydrodynamical approximation request ($\lambda\sim L\gg\xi$) and for them the analogy works. \footnote{This is more than a guess, since it is partially confermed by the results in this direction coming from the Casimir effect \cite{kempf}, even if so far there is not a conclusive answer.}\\ Using techniques developed within QFT in curved space, it is possible to overtake the difficulties in handling analytical calculations for non-homogeneous BEC \cite{bec}. Now, noting that 
 \be\tilde \rho_{in}=\langle \hat \varphi^\dagger\hat\varphi\rangle=\frac{\langle
\hat \rho_1^2\rangle}{4\rho}+\frac{\rho \langle \hat \theta_1^2\rangle}{\hbar^2}
\approx \frac{\rho \langle \hat
\theta_1^2\rangle}{\hbar^2}\left(1+O\left(\frac{\xi^2}{L^2}\right)\right)\, , \label{nin}\ee
the calculation of the depletion has been reduced to the calculation of the expectation value of the squared of a massless scalar field minimally coupled to the acoustic spacetime (\ref{gmunu}).
For a spherically symmetric static BEC configuration, using renormalization techniques peculiar of QFT in curved space, it is possible to give an explicit description for the thermal and quantum depletion for an inhomogeneous BEC in term of a general density $\rho(r)$:
\bea
\label{our.dep}
&&\tilde \rho_{in}
=\frac{\kappa_B^2 T^2}{12\hbar^3 }\left(\frac{m^3}{ g\rho}\right)^{1/2}-\frac{(mg\rho)^{1/2}}{96\pi^2 \hbar }\left[-\frac{7}{8} \frac{\rho'^{2}}{\rho^2}+\frac{5}{2}\frac{\rho''}{\rho}+\right.\nonumber\\ &&\left.+\frac{5}{r}\frac{\rho'}{\rho}\right]\ln \mu^2\frac{g^{1/2} \rho^{3/2}}{m^{3/2}}+\frac{(mg\rho)^{1/2}}{48\pi^2 \hbar }\left[ -\frac{39}{16}\frac{\rho'^{2}}{\rho^2}+\frac{3}{4}\frac{\rho''}{\rho}+\frac{3}{2r}\frac{\rho'}{\rho}
\right]\, ,
\eea
where $m$ is the mass of the atoms and $g$ is a constant proportional to the scattering length $a$: $g=4\pi\hbar^2 a/m$.
Note that the thermal term exactly reproduces   the thermal depletion in LDA.  The other terms instead generalize the zero-temperature depletion in LDA taking the spatial derivatives of the condensate density explicitely into account. In fact  all of them vanish for a homogeneous BEC.\\
Moving to configurations with a radial stationary velocity ($\vec v=v(r)\vec r/|\vec r|$), the concept of thermal equilibrium loses its meaning, but it is still possible to evaluate the effects of this velocity on the quantum depletion, yielding \cite{bec}
\bea \label{our.dep.v} &&\tilde
n_{in}^q=- \frac{m}{96\pi^2 \hbar c^2}\left[
\frac{3}{2}cc'^2+5c^2c''+-3v^2c''-2cv'^2-2cvv''
+\frac{9}{2c}v^2 c'^2\right.\nonumber\\ &&\left.-4vc'v'+\frac{2}{r}(5c^2c'-4cvv'-v^2c')
-\frac{2}{r^2}v^2c\right]\ln \mu^2\frac{c(c^2-v^2)}{g}+\nonumber\\
&&+\frac{m}{48\pi^2 \hbar c^2}\left\{ -\frac{1}{2c(c^2-v^2)}(3c^2c'-v^2c'-2cvv')(3c^2c'-2v^2c'-cvv')\right.\nonumber\\ &&\left.-\frac{15}{4}cc'^2-\frac{1}{4c}v^2c'^2-\frac{5}{2}vc'v'++\frac{3}{2}c^2c''-\frac{1}{2}v^2 c''-cv'^2-cvv''+\right.\nonumber\\ &&\left.+\frac{1}{r}(3c^2c'-v^2c'-2cvv')\right\}\,  ,
\eea
where we have introduced the local sound speed $c$, that for BEC is related to the density by the equation of state $mc^2=gn$.
This expression generalizes eq.(\ref{our.dep}) considering also contributions due to the velocity of the condensate. It reduces to eq.(\ref{our.dep}) only for $v(r)=0$: for $v=const$ in fact a curvature is induced since a radial velocity can not be trivially eliminated by a Galileo transformation.\\
In the case of an acoustic black hole, the expression in eq.(\ref{our.dep.v}) diverges on the acoustic horizon ($v^2=c^2$): this is a familiar divergence in GR, related to the fact that the quantum state over which the expectation value of $\langle \hat \theta_1^2\rangle$ has been calculated is not the right one to describe  the region next to the horizon. In GR it has been shown that for a Schwarzschild  black hole the same calculation for the Unruh state leads to a regular result: the same is expected also for acoustic black holes.

\section*{Conclusions}
In this paper I have illustrated the first step in extending the analogy between gravitation and condensed matter systems also to dynamical issues. It has been stressed how this extension has to be handled with particular care, without expecting in the acoustic setting  an exact counterpart  of the results already obtained in GR (unlike what happens with the kinematical ones).\\
However there is still a crucial issue to be addressed: how big are the errors in neglecting high momenta effects? \\ This question involves both QFT in curved space and condensed matter physics. As already mentioned, for BEC as in general for fluids, very high $k$ modes see the coarse grain of the medium, and, as a consequence, they do not obey the relativistic dispersion relation, but a short-distance modified one.\\ In the gravitational context instead the question is  related  to the so called {\it transplanckian  problem}: the frequency of the modes which give the Hawking radiation is indeed transplanckian, i.e. beyond the actual reliability limit of the theory used to study the process, namely QFT in curved space.
This fact does raise doubts on the reliability of the Hawking effect itself, as a consequence of the breaking of QFT in curved space beyond Planck energy scale, where Quantum Gravity effects should become important. Thought a full Quantum Gravity theory is not known and therefore it is not possible to safely say how these behaviors will emerge, it is widely believed that a first hint of such effects should show up in a short-distance modification of the dispersion relation \cite{mattingly} (like what happens  in the acoustic setting). Recent investigations on Hawking effect within a QFT with modified dispersion relation have shown that Hawking radiation is only slightly affected by these modifications, turning out to be unaltered in its fundamental features \cite{unruh-sch, review}.\\
This is a good starting point to carry out deeper investigations involving also the calculation of expectation values in presence of such modified dispersion relations. Moreover also the first result concerning the calculation of renormalized expectation value goes in this direction: in fact the Casimir energy density between two conducting plates can be evaluated with great accuracy using the phonon dispersion relation, provided the plate separation is much larger than the interatomic distance of the constituents of the plates \cite{kempf}. Similar analysis has been performed also  in the cosmological context,  yielding so far only partial results \cite{mazzitelli}. Nothing has still been obtained for a black hole background.\\ A deeper knowledge of these effects will be of crucial importance in several field: cosmology, black hole physics and cold-atoms systems. A more accurate use of QFT in curved space will turn out to be very useful both in studying gravitation directly and in leading a more reliable use of the analogy in the dynamics of the condensed matter systems. \\

{\bf Acknowledgments}: I would like to thank R. Balbinot for helpful suggestions and contributions to this paper. I also thank "Enrico Fermi" Center for supporting this research.

\end{document}